\documentclass[10pt]{iopart}
\usepackage{graphicx,epsfig}
\usepackage{bm}
\usepackage{iopams}
\usepackage{color,ulem}
\newcommand\eqref[1]{(\ref{#1})}
\newcommand\etall{{\it {et al}}}
\newcommand{\ba}{\begin{eqnarray}}
\newcommand{\ea}{\end{eqnarray}}

\newcommand{\B}{{\cal{B}}}
\newcommand{\M}{{\cal {M}}}

\newcommand{\bbq}{\begin{quote}}
\newcommand{\eeq}{\end{quote}}

\newcommand{\tbb}{t_{\textrm{\tiny{bb}}}}

\newcommand{\tcoll}{t_{\textrm{\tiny{coll}}}}
\newcommand{\tmax}{t_{\textrm{\tiny{max}}}}

\newcommand{\FF}{{\cal{F}}}

\newcommand{\MM}{{\cal{M}}}

\newcommand{\dd}{{\rm{d}}}

\usepackage{pdfsync}

\def\Journal#1#2#3#4{{#1} {\bf #2}, #3 (#4)}
\def\AA{{Astron. Astrophys.}}

\def\AJ{{Astron. J.}}

\def\APJ{{Astrophys. J.}}

\def\CQG{{Class. Quant. Grav.}}
\def\EPJC{{Eur. Phys. J. C}}
\def\GRG{{Gen. Relativ. Gravit.}}

\def\JCAP{{JCAP}}

\def\LRR{{Living Rev. Rel.}}

\def\PLB{{Phys. Lett.}  B}

\def\PRL{Phys. Rev. Lett.}

\def\PRD{{Phys. Rev.} D}
\def\PR{{Phys. Rep.}}

\def\RMP{{Rev. Mod. Phys.}}

\begin{document}

\title[Lema\^\i tre--Tolman--Bondi dust solutions in $f(R)$ gravity.]{Lema\^\i tre--Tolman--Bondi dust solutions in $f(R)$ gravity.} 
\author{ Roberto A. Sussman and Luisa G. Jaime}
\address{Instituto de Ciencias Nucleares, Universidad Nacional Aut\'onoma de M\'exico (ICN-UNAM), A. P. 70-543, 04510, CDMX, M\'exico }
\eads{$^\ddagger$\mailto{sussman@nucleares.unam.mx}}
\date{\today}
\begin{abstract} 
We derive a class of non--static inhomogeneous dust solutions in $f(R)$ gravity described by the Lema\^\i tre--Tolman--Bondi (LTB) metric. The field equations are fully integrated for all parameter subcases and compared with analogous subcases of LTB dust solutions of GR. Since the solutions do not admit regular symmetry centres, we have two possibilities: (i) a spherical dust cloud with angle deficit acting as the source of a vacuum Schwarzschild--like solution associated with a global monopole, or (ii) fully regular dust wormholes without angle deficit, whose rest frames are homeomorphic  to the Schwarzschild--Kruskal manifold or to a 3d torus. The compatibility between the LTB metric and generic  $f(R)$ ansatzes furnishes an ``inverse procedure'' to generate LTB solutions whose sources are found from the $f(R)$ geometry. While the resulting fluids may have an elusive physical interpretation, they can be used as exact non--perturbative toy models in theoretical and cosmological applications of $f(R)$ theories.                                                        
\end{abstract}

\maketitle
\section{Introduction.} 

The concordance $\Lambda$CDM model provides a well known fit to a wide range of different independent observations \cite{SNIa,Planck2015-Overview,Planck2015-CosmoParam,Sahni2014}. However, the detection of dark matter and finding a consistent theoretical support for dark energy and/or the cosmological constant (which enable for an accelerated cosmic expansion) have proven to be an elusive and difficult task. Hence, alternative gravity theories have been proposed to produce the accelerating mechanism through extra geometric degrees of freedom without assuming dark sources or a cosmological constant \cite{Lombriser2016,Nojiri2006,Nojiri2011}. Perhaps the most straightforward modification of the Einstein equations are $f(R)$ theories of gravity (for a review see \cite{Sotiriou2010,Jaime2012,deFelice2010}). These theories of gravity have been used to produce acceleration of the Universe at different stages. In particular, the most successful  models compatible with observations at different scales are those proposed  by Starobinsky \cite{Starobinsky2007} and by Hu and Sawicki \cite{Hu2007}. 

Most work in $f(R)$ theories involve homogeneous and isotropic Friedmann--Lema\^\i tre--Robertson--Walker (FLRW) metrics in a cosmological context (see review in \cite{Jaime2012} and \cite{luisa1}), but they have been used also in an inflationary scenario \cite{Rinflation}. Exact solutions are known for vacuum and black hole sources \cite{Canate2016, Nzioki2016, Clifton2006, Sebastiani2011, Habib2012}, as well as static spherically symmetric sources \cite{Babichev2009, Upadhye2009, Kobayashi2008, alvaro, Jaime2011, Astashenok2017}. More recently various articles have studied the spherical gravitational collapse, either numerically \cite{borisov} or through exact  solutions \cite{chakrabarti, goswami, cembranos} involving inhomogeneous non--static sources that can be described as mathematical fluids which, most likely, lack a proper physical interpretation (though an interesting qualitative study providing useful physical guidelines for ``tilted'' LTB metrics is found in \cite{yousaf}).    

In the present paper we examine the compatibility between $f(R)$ theories and the well known spherically symmetric Lema\^\i tre--Tolman--Bondi (LTB) metric. Within the framework of standard General Relativity (GR) this metric is associated with exact solutions with a dust source \cite{kras1,kras2,BKHC,ellis} (though nonzero pressure is possible \cite{kras1,suss2009}), which provide valuable toy models that have been widely used, not only in cosmological applications \cite{BCK,marra1,marra2,biswas,suss1}, but also to explore open theoretical issues \cite{entropy,suss2} under a mathematically tractable approach that still allows for a full non--perturbative description of non--linear effects. Hence, it is reasonable to expect that similar useful LTB toy models could be devised for $f(R)$ theories.  

The $f(R)$ field equations simplify considerably for the LTB metric, since all derivatives of $f(R)$ higher than first order can be eliminated in terms of (up to second order) time and radial coordinate derivatives of $f_R=\dd f(R)/\dd R$ and of the main metric function: the area distance $Y$. Also, the Ricci scalar ($R$) itself directly yields the Friedman--like equation that governs the time evolution of comoving observers, and thus we are able to integrate the $f(R)$ field equations up to a first order quadrature for all $f(R)$ ansatzes for which the derivative $f_R(R)$ is an invertible function that yields $R=R(Y)$ (see section \ref{sec:field-equations}). This approach allows us to generate an ``inverse'' procedure (similar to that of \cite{Nojiri2011} for FLRW metrics) in which a fluid source (with anisotropic pressure) for the LTB metric is determined by the geometry through the choice of $f(R)$ and the corresponding field equations (section \ref{solving}). Evidently, the sources generated by such a procedure are most likely mathematical fluids that can be hard to interpret physically (we discuss this issue in section \ref{pressure}).  

To avoid dealing with mathematical fluids we show in section \ref{dustsols} that a simple choice $f(R)\propto \sqrt{R}$ exists that yields a class of exact solutions whose source is physically meaningful: dust with an effective cosmological constant (which can always be set to zero). The field equations are fully integrated for all parameter sub--cases (see \ref{append}). As far as we are aware, this is the first example in the literature of an exact non--static and inhomogeneous class of dust solutions in $f(R)$ gravity. 

Having obtained this class of exact dust solutions we examine their physical and geometric properties. Two noticeable features readily emerge: (i) the solutions do not admit a regular symmetry centre (section \ref{deficit}) and (ii) their vacuum sub--case is the Schwarzschild--like spacetime with angle deficit found and discussed in detail in \cite{Canate2016,Habib2012}, which is not asymptotically flat (section \ref{vacuum}). Hence, if used as models of localised sources with a vacuum exterior these dust solutions {\it necessarily} describe a spherical dust cloud with a global monopole at its centre, acting as the source of the vacuum solution of \cite{Canate2016,Habib2012}. As we prove in section \ref{matching}, this vacuum exterior can be smoothly matched (generically) to the dust cloud interior. However, this is not the only possibility: as we argue in section \ref{wormholes}, perfectly regular dust solutions without angle deficit are possible if we do not demand the existence of a regular centre and/or a smooth matching with a vacuum sub--case. Such solutions are effectively dust wormholes \cite{hellaby,matravers,humphreys} whose rest frames (hypersurfaces of constant $t$ orthogonal to the 4--velocity) are homeomorphic to ${\bf S}^2\times {\bf R}$ or to a 3--d torus ${\bf S}^2\times {\bf S}^1$.    

We examine in section \ref{pressure} the resulting sources for the $f(R)$ ansatz proposed by Miranda {\it et al} \cite{miranda} in an FLRW cosmological context, which also allows for a functional form of $f_R$ that is readily invertible. As expected, the source obtained through the inverse procedure for this $f(R)$ corresponds to a mathematical fluid with various possible physical pathologies, but that nevertheless reproduces in the asymptotic time range the accelerated expansion expected from observations fitting. We also examine and comment on the inhomogeneous non--static solutions found in the literature, all of which can be derived from the inverse procedure we have discussed as particular cases. Finally, we provide a brief summary and conclusions in section \ref{final}. The analytic solutions of the Friedman--like evolution equation for dust layers are given in full in \ref{append}.

\section{Field equations}
\label{sec:field-equations}

The field equations for $f(R)$ theories are given by
\ba f_{R}G_{ab}-f_{RR}\nabla_a\nabla_b\,R-f_{RRR}\nabla_a R\nabla_b R\nonumber\\
 +\left[\frac{R\,f_{R}-f}{2}+f_{RR}\Box R+f_{RRR}\nabla_c\nabla^c R\right]g_{ab}=8\pi T_{ab},\label{feqs}
\ea
where $R$ is the Ricci scalar and we are using the notation $f_{R..R}=\dd^n f/\dd R^n$ for the n--th derivative of $f(R)$, while $\Box R = g^{cd}\nabla_c\nabla_d R$ and $\nabla^c R\nabla_c R= g^{cd}\nabla_c R\nabla_c R$. 

We consider the fully general spherically symmetric Lema\^\i tre--Tolman--Bondi (LTB) metric element
\begin{equation} \dd s^2 = -\dd t^2 + \frac{Y'^2}{1-K}\dd r^2+Y^2\left(\dd\theta^2+\sin^2\theta\,\dd\phi^2\right),\label{LTB}\end{equation}    
where $Y=Y(t,r),\,\,K=K(r)$ and $Y'=\partial Y/\partial r$ (prime will denote derivative with respect to $r$). The Ricci scalar, $R$, associated with (\ref{LTB}) is 
\begin{equation} R\,Y^2\,Y' =\left[2Y^3\left(\frac{\ddot Y}{Y}+\frac{\dot Y^2+K}{Y^2}\right)\right]',\label{Ricci}\end{equation}
where $\dot Y=\partial Y/\partial t$ (dot will denote derivative with respect to $t$). We consider a matter source in a comoving frame whose energy momentum tensor is
\begin{equation} T_{ab}=\rho\,u_a u_b+p\,h_{ab}+\Pi_{ab},\label{Tab}\end{equation}
where $u^a=\delta^a_t$,\,\,
 $h_{ab}=u_a u_b+g_{ab}$, while $\rho(t,r)$ and $p(t,r)$ are the energy density and isotropic pressure and the symmetric traceless anisotropic pressure tensor is given by  $\Pi^a_b=P(t,r)\,\hbox{diag}[0,-2,1,1]$.
The field equations (\ref{feqs}) for the metric (\ref{LTB}) take the form
\ba \fl 0 = \dot f'_R-f'_R\frac{\dot Y'}{Y'}\label{Feq41}\\
\fl\frac{f_R\left[Y(\dot Y^2+K)\right]'}{Y^2Y'}=8\pi\rho
+\frac{1}{2}\left(f_R\,R-f\right)+\frac{(1-K)f''_{R}}{Y'^2}-\left(\frac{\dot Y'}{Y}+\frac{2\dot Y}{Y}\right)\dot f_{R}\nonumber\\
\fl -\frac{\left[2(1-K)(YY''+2Y'^2)+YY'K'\right]f'_{R}}{3YY'^3},\label{Feqrho}\\
\fl -\frac{f_{R}\left[2Y^2Y''-Y(\dot Y^2+K)\right]'}{3Y^2Y'}=8\pi p-\frac{1}{2}\left(f_{R}R-f\right)+\ddot f_{R}-\frac{2(1-K)f''_{R}}{3Y'^2}
\nonumber\\
\fl+\frac23\left(\frac{\dot Y'}{Y}+\frac{2\dot Y}{Y}\right)\dot f_{R}+\frac{\left[2(1-K)(YY''-2Y'^2)+YY'K'\right]f'_{R}}{3YY'^3},\label{Feqp}\\
\fl -\frac{f_{R}Y}{6Y'}\left[\frac{2\ddot Y}{Y}+\frac{\dot Y^2+K}{Y^2}\right]'=8\pi P-\frac{(1-K)f''_{R}}{3Y'^2}+\frac13\left(\frac{\dot Y'}{Y}-\frac{\dot Y}{Y}\right)\dot f_{R}\nonumber\\
\fl+\frac{\left[2(1-K)(YY''+Y'^2)+YY'K'\right]f'_{R}}{6YY'^3}.\label{FeqPP}\ea 
The GR limit follows readily from $f(R)=R$, so that $f_{R}=1$ (notice that the $[tr]$ field equation (\ref{Feq41}) is trivial, since $G^t_r=G^r_t=0$ holds for the LTB metric).  

\section{Solving the field equations.}\label{solving}

The solution of the off--diagonal field equation (\ref{Feq41}) follows readily as
\begin{equation} f_{R}=\beta_0+\frac{Y}{\ell_0(r)}-\int{Y\,\left(\frac{1}{\ell_0}\right)'\dd r},\label{f1}\end{equation}
where the function $\ell_0$ has length units and $\beta_0$ is an arbitrary dimensionless constant. We shall assume henceforth that $\ell'_0=0$, so that $\ell_0$ marks a constant characteristic length scale. Inserting (\ref{Ricci}) and (\ref{f1}) into (\ref{Feqrho})--(\ref{FeqPP}) and rearranging terms, equation (\ref{f1}) leads to
\ba \fl 8\pi\rho = \frac{1}{2}f-\frac{(Y/\ell_0+\beta_0)\left(Y^2\ddot Y\right)'}{Y^2Y'}+\frac{\left[Y^4(\dot Y^2+K-1)\right]'}{2\ell_0Y^4Y'},\label{eqrho}\\
\fl 8\pi p = -\frac{1}{2}f+\frac{(Y/\ell_0+\beta_0)\left[Y^2\ddot Y+2Y(\dot Y^2+K)\right]'}{3Y^2Y'}-\frac{\left[Y^4(\dot Y^2+K-1)\right]'}{3\ell_0Y^4Y'}-\frac{\ddot Y}{\ell_0},\nonumber\\
\label{eqp}\\
\fl 8\pi P = -\frac{Y}{6Y'}\left[\left(\frac{Y}{\ell_0}+\beta_0\right)\left(\frac{2\ddot Y}{Y}+\frac{\dot Y^2+K}{Y^2}\right)'+\frac{Y}{\ell_0}\left(\frac{\dot Y^2+K-1}{Y^2}\right)'\right],\label{eqPP}
\ea 
where the GR limit is recovered by setting $f(R)=R$ (with $R$ given by (\ref{Ricci})) together with $\beta_0=1$ and $Y/\ell_0\to 0$ (so that $f_{R}=1$). In fact, $\ell_0$ can be selected for a given $f(R)$ to mark the length scales (in terms of $Y$) at which GR is recovered. 

Since $f_{R}=f_{R}(R)$, in principle we can obtain the Ricci scalar in terms of the functional inverse $f_{R}^{-1}$:
\begin{equation} R = R(Y) \equiv f_{R}^{-1}(Y/\ell_0+\beta_0),\label{inverse}\end{equation}
leading (in view of the expression for $R$ in (\ref{Ricci})) to the following Friedman--like equation
\begin{equation}\frac{\ddot Y}{Y}+\frac{\dot Y^2+K}{Y^2}=\frac{1}{6} R_q,\label{Ricci3}\end{equation}
with $R_q$ defined as the following integral of the Ricci scalar obtained as a function of $Y$ from (\ref{inverse})
\begin{equation}\fl R_q = \frac{3}{Y^3}\int_{r_*}^r{R(Y)\,Y^2\,Y'\,\dd r}\quad \Rightarrow\quad R=R_q+\frac{R'_q}{3Y'/Y},\label{Rqdef}\end{equation}
where the lower limit in the integral above can always be fixed by boundary conditions.  

The Friedman--like equation (\ref{Ricci3}) is, in principle, integrable for those $f(R)$ in which the functional inverse $R=f_{R}^{-1}$ in (\ref{inverse}) has a closed analytic form. For a given $f(R)$ the solutions of (\ref{Ricci3}) allow us to set up an ``inverse'' procedure to determine $\rho,\,p$ and $P$ from (\ref{eqrho})--(\ref{eqPP}).

A first integral of (\ref{Ricci3}) follows readily if we introduce the change of variable $Y=\sqrt{X}$, leading to
\begin{equation} \dot Y^2 = -K +\frac{\MM}{Y^2},\qquad \frac{\ddot Y}{Y}=-\frac{\MM}{Y^4}+\frac16 R_q,\label{dotYsq}\end{equation}
where $\MM=\MM(t,r)$ is given by 
\begin{equation}\fl\MM\equiv \mu(r)\ell_0+\frac{1}{3}\int_t{R_q Y^3\dot Y\,\dd t} = \mu(r)\ell_0+\int{\dd t\,\dot Y\int_{r_*}^r{\dd r\,R(Y)Y^2Y'}}.\label{MMdef}\end{equation}
and $\mu(r)$ emerges as an integration constant with length units (so that $\mu\ell_0$ is a squared length).
Substitution of (\ref{dotYsq}) and (\ref{MMdef}) into (\ref{eqrho})--(\ref{eqPP}) yields
\ba \fl 8\pi\rho = \frac{f}{2}+\frac{Y/\ell_0+\beta_0}{6Y^2Y'}\left(\frac{6\MM}{Y}-R_qY^3\right)'+\frac{\left[Y^2(\MM-Y^2)\right]'}{2\ell_0Y^4Y'},\label{eqrho2}\\
\fl 8\pi p = -\frac{f}{2}+\frac{1}{6\ell_0Y^2}\left(\frac{6\MM}{Y}-R_qY^3\right)+\frac{Y/\ell_0+\beta_0}{18Y^2Y'}\left(\frac{6\MM}{Y}+R_qY^3\right)'-\frac{\left[Y^2(\MM-Y^2)\right]'}{3\ell_0Y^4Y'},\nonumber\\
\fl\label{eqp2}\\
\fl 8\pi P = -\frac{Y}{18Y'}\left[\frac{Y}{\ell_0}\left(R_q-\frac{3}{Y^2}\right)'+\beta_0\left(R_q-\frac{3\MM}{Y^4}\right)'\right],\label{eqPP2}\ea
Exact solutions follow from (\ref{eqrho2})--(\ref{eqPP2}) for any given choice of $f(R)$ where the  functional inverse $R=R(Y)=f_{R}^{-1}(Y)$ is expressible in terms of closed analytic forms. 

\section{A class of dust solutions.}\label{dustsols}

The condition for a perfect fluid solution follows by setting $P=0$ in (\ref{eqPP2}), leading to
\begin{equation} \frac{Y}{\ell_0}\left(R_q-\frac{3}{Y^2}\right)'+\beta_0\left(R_q-\frac{3\MM}{Y^4}\right)'=0.\label{PFcond}
\end{equation}
The following simple class of particular solutions of \eqref{PFcond} is obtained  by assuming that $\beta_0=0$ (notice that we already assume that $\ell_0$ is constant in \eqref{f1}):
\begin{equation}\fl R_q(Y) = \frac{3}{Y^2}+R_*\qquad\Rightarrow\qquad  R(Y)=\frac{1}{Y^2}+R_*,\label{PF1}\end{equation}
where the $R_*$ is an integration constant with inverse square length units.  

From (\ref{f1}), (\ref{inverse}) and (\ref{MMdef}) we obtain $f_{R},\,f$ and $\MM$, either as functions of $R$ or as functions of $Y$
\ba f_{R}(R)=\frac{\dd f}{\dd R}=\frac{1}{\ell_0\sqrt{R-R_*}}=\frac{Y}{\ell_0},\label{PFf1}\\
 f(R)=\lambda+\frac{2\sqrt{R-R_*}}{\ell_0}=\lambda+\frac{2}{\ell_0Y},\label{PFf}\\
 \MM = \mu(r)\ell_0+\frac12 Y^2+\frac{R_*}{12}\,Y^4,\label{PF3}\ea
where $\lambda$ is a second dimensionless arbitrary constant (also an inverse squared length) that is independent of $R_*$. Inserting (\ref{PF1})--(\ref{PF3}) into (\ref{dotYsq}),\, (\ref{eqrho2}) and (\ref{eqp2}) we obtain after some algebraic manipulation the following forms for the energy density and pressure of the source 
\begin{equation} 8\pi\rho = \frac{\frac32 \mu'}{Y^2Y'}+\frac{\lambda}{2},\qquad 8\pi p = -\frac{\lambda}{2}.\label{exact1}\end{equation}
The kinematic evolution of this dust source is governed by the Friedman--like equation (\ref{dotYsq}), which upon substitution of \eqref{PF3} takes the form
\begin{equation}\dot Y^2 = \frac12-K+\frac{\mu\ell_0}{Y^2}+R_* Y^2,\qquad \frac{\ddot Y}{Y}=-\frac{\mu\ell_0}{Y^4}+R_*.\label{exact2}\end{equation}
Although the exact solutions (\ref{exact1})--(\ref{exact2}) look qualitatively similar to the LTB solution for dust and a cosmological constant of GR, they have distinct physical and geometric properties, which we discuss in detail below and in forthcoming sections. 

The first striking difference with respect to GR solutions is the fact that we obtain two ``effective'' cosmological constants, $\lambda$ and $R_*$, which are independent of each other: $\lambda$ in (\ref{PFf}) emerges when integrating $f_R$ to obtain $f(R)$, and thus $\lambda>0$ implies a choice of $f(R)$ that already incorporates a cosmological constant in the Einstein--Hilbert action (the exact identification is $\lambda=16\pi\Lambda/3$), whereas $R_*$ follows from the perfect fluid condition $P=0$ in (\ref{PFcond}), irrespective of the choice of $f(R)$. As a consequence, we have the following interesting possibilities:
\begin{itemize}
\item If we choose $\lambda=0$ and $R_*>0$, a density in \eqref{exact1} without cosmological constant (and $p=0$) produces a large time de Sitter kinematic evolution associated with a $\Lambda$ term given by $R_*$ ({\it i.e.} $\dot Y/Y\sim \sqrt{R_*}$ for large $Y$).
\item If we choose $\lambda>0$ and $R_*=0$ we obtain the opposite effect:  density and pressure with cosmological constant in \eqref{exact1} produce a kinematic evolution without a $\Lambda$ term. 
\item If we choose $\lambda=R_*>0$, we obtain the GR behaviour: an effective cosmological constant in $\rho$ and $p$ produces the kinematic evolution associated with a $\Lambda$ term. We will assume henceforth that $\lambda=R_*>0$. 
\end{itemize}
However, notice that the obtained solutions are still consistent if we assume that $\lambda$ and $R_*$ are nonzero but different, with same or opposite sign, or if both are set to zero.  The following are other important features worth highlighting 
\begin{itemize}
\item The kinematic behaviour of dust layers is similar to their GR analogues, and for all choices of these functions the analytic solutions follow from the integral quadrature of the equation for $\dot Y^2$ in (\ref{exact2}). We present all these analytical solutions in the Appendix.
\item The function $\M$ is apparently analogous to the Misner--Sharp mass function in GR, but this analogy must be handled carefully: $\M$ does not satisfy the precise properties that characterise this function in  spherically symmetric GR solutions for the fluid source \eqref{Tab}: the integrability conditions that furnish relations between its time/radial derivatives and the density and pressures: $\M'\ne 4\pi\rho Y^2Y'$ and $\dot\M\ne -4\pi (p-2P) Y^2\dot Y$.  
\item The LTB dust solutions we have derived (see \ref{append}) admit two types of singularities for which $\rho$ and curvature scalars diverge: (i) a Big Bang or (for re--collapsing models) Big Crunch and (ii) Shell crossings, respectively marked by $Y(t,r)>0$ and $Y'(t,r)=0$ for $r\ne$ constant. As with LTB dust solutions in GR, the singularities (i) are unavoidable but the free parameters can always be selected so that shell crossings (ii) are avoided. However, as we find out in the following section, the LTB solutions \eqref{sol1}--\eqref{sol8} may also present a conical singularity at $r=0$ that is absent in their GR counterparts. 
\end{itemize}

\section{Solutions with angle deficit and conical singularity.}\label{deficit} 

A regular symmetry centre is the worldline (marked by $r=0$) of a fixed point of SO(3), defined by  the following boundary conditions   
\begin{equation}\fl Y(t,0)=\dot Y(t,0)=0,\quad Y'(t,0)=a(t)>0,\quad K(0)=0,\quad \mu(0)=0,\label{centre}\end{equation}
which must hold for all $t$ and for finite curvature scalars. Conditions \eqref{centre} imply considering $r\geq 0$ as the radial range the integral $\int{RY^2Y'\dd r}$ in \eqref{Rqdef}. A regular symmetry centre requires also the following conditions at first order for $r\approx 0$ \cite{kras2,matravers,humphreys} 
\begin{equation} \fl Y\sim \bar a(t)\,r+O(r^2),\qquad K\sim O(r^2)\qquad \mu\sim r^n,\quad n\geq 3.\label{centre2}\end{equation}
Evidently, the regular centre conditions (\ref{centre}) and (\ref{centre2}) do not hold for solutions (\ref{exact1})--(\ref{exact2}). While the dust density obtained from \eqref{exact1} need not diverge in the limit $Y\to 0$ as $r\to 0$,  the Ricci scalar in \eqref{PF1} necessarily diverges in this limit for whatever choice of free functions $K,\,\mu,\,\tbb$ (see \ref{append}). In fact, the same singular behaviour occurs at $r=0$ if we demand fulfilment of \eqref{centre}--\eqref{centre2} for all other polynomial curvature scalars, as for example the Kretschmann scalar
\begin{equation}\fl  R_{abcd}R^{abcd}= \frac{1}{Y^4}+6\lambda^2+\frac{2\lambda}{Y^2}+\frac{4\mu}{Y^6}\left(1-\frac{8\mu'}{YY'}\right)+\frac{6\mu'^2}{Y^6Y'}.\end{equation}
where we notice that only in those terms where $\mu'$ appears in the numerator a zero of $Y$ at $r=0$ can be removed but this is not possible if the numerator is a constant.   

To explore the lack of a regular centre we remark that, besides the blowing up of curvature scalars at $r=0$, the term $1/2$ in the right hand side of the Friedman--like equation (\ref{exact2}) suggests the presence of a conical singularity at $r=0$ in the equatorial surface $\theta=\pi/2$, which signals  an azimuthal angle deficit $\pi/2$ that can be associated with topological defects, such as global monopoles or cosmic strings \cite{barriola,nucamendi} (see also \cite{Canate2016,Habib2012,tahamtan}). A regular centre requires local flatness around $r=0$, which in turns requires $K(0)=0$ and $\mu(0)=0$. While we can always choose $K$ and $\mu$ to comply with these conditions, local flatness does not hold: by rescaling the radial coordinate as $\bar r=\sqrt{2}r$ the infinitesimal spherical world tube region around $r=0$ is described by the metric $\dd s^2=-\dd t^2+a^2(t)[\dd\bar r^2+(1/2)\bar r^2(\dd\theta^2+\sin^2\theta\dd\phi^2)]$, which is not the Minkowski metric (the surface area of such an infinitesimal comoving sphere is not $4\pi\bar r^2\bar a^2(t)$ but $2\pi\bar r^2\bar a^2(t)$). This lack of local flatness around $r=0$ signals an angle deficit and the existence of a conical singularity as generic features that characterise the whole class of dust solutions (\ref{exact1})--(\ref{exact2}) when the centre conditions \eqref{centre}--\eqref{centre2} are imposed.
 

Since the centre conditions \eqref{centre}--\eqref{centre2} are not incompatible with a bounded dust density as $r\to 0$, we could consider the angle deficit and its associated conical singularity as features of the $f(R)$ geometry, or as indicative of a distributional source associated with a topological defect: a cosmic string or a global monopole \cite{barriola,nucasud,nucamendi,Canate2016}. In the latter case we would have to add to \eqref{exact1} a distributional contribution to the density in \eqref{exact1}: $8\pi[\rho+\eta^2\delta(r)]$ accounting for the global monopole at the centre $r=0$, where $\eta$ is the length scale of the global symmetry breaking (see detailed explanation in \cite{barriola,nucasud}). While this distributional source would contribute to the mass--energy obtained from an integral analogous to the Misner--Sharp mass, as argued in \cite{barriola}, such contribution is negligible in macroscopic astrophysical scales. Therefore, as long as we consider non--trivial dust configurations with $\mu'\ne 0$ and $\rho>0$ we will henceforth neglect the contribution from this type of distributional source. Considering (for simplicity) $\lambda=0$ in \eqref{exact1}, the range $r>0$ for the radial integral and the angle deficit $\Delta=1/2$ in the periodicity range $0\leq \phi<\pi$ for the integral over $\phi$, we obtain the following mass integral that is analogous (save the angle deficit factor) to the Misner--Sharp mass $M$ of GR
\begin{equation}\fl \frac{G}{c^2}
\int\limits_{0}^\pi\,\dd\phi\! \int\limits_{0}^\pi\dd\theta\,\sin^2\theta\!\int\limits_{0}^r{
    {\dd r\,\rho Y^2Y'}
             } =\frac{2\pi G}{c^2}\int_0^r{\rho Y^2Y'\dd r} =\frac{3\mu(r)}{2},  \label{massInt}\end{equation}
where we remark that we are omitting the negligible \cite{barriola}  but (strictly speaking) nonzero mass--energy contribution associated with the monopole charge.    
%

\section{Vacuum subcase with angle deficit}\label{vacuum}

The subcase in which $\mu'=0$ holds everywhere yields from (\ref{exact1})--(\ref{exact2}) a non--trivial $f(R)$ locally vacuum solution ($\rho=\lambda/2=-p$) characterised by an angle deficit and a conical singularity at $r=0$. In the absence of the dust source this solution is an example of a spacetime that is ``asymptotically flat save for the angle deficit'' \cite{nucasud,nucamendi}, whose source can be a global monopole.  

Exact spherically symmetric solutions of this type in $f(R)$ geometry have been examined before and are always static \cite{Canate2016,Habib2012,barriola,tahamtan}. However, the exact solution that follows from \eqref{exact1}--\eqref{exact2} with $\mu'=0$ is also static, even if described by the non--static LTB metric \eqref{LTB}. In fact, it is straightforward to show that the coordinate transformation $t=t(T,Y),\,\,r=r(T,Y)$ given by
\ba \fl Y'\dd r =\dd Y -\dot Y\frac{\partial t}{\partial T}\dd T,\qquad \dd t = \frac{\partial t}{\partial Y}\dd Y+\frac{\partial t}{\partial T}\dd T,\nonumber\\
\fl \frac{\partial t}{\partial Y} = \frac{\sqrt{\frac{\mu_0\ell_0}{Y^2}-K+\frac12+\frac{\lambda}{2}Y^2}}{\frac12\left(1-\frac{\mu_0\ell_0}{Y^2}+\frac{\lambda}{2}Y^2\right)},\qquad \frac{\partial t}{\partial T} =\sqrt{1-K},\nonumber
\ea
brings the LTB metric (\ref{LTB}) for this vacuum solution to the very recognisable Schwarzschild--like or Kottler--like line element in a static frame
\ba\fl\dd s^2 = -\frac12\left(1-\frac{\mu_0\ell_0}{Y^2}+\frac{\lambda}{2}Y^2\right)\dd T^2+\frac{\dd Y^2}{\frac12\left(1-\frac{\mu_0\ell_0}{Y^2}+\frac{\lambda}{2}Y^2\right)}+Y^2\left(\dd\theta^2+\sin^2\theta\dd\phi^2\right).\nonumber\\
\label{Schw}\ea
where we can identify $\mu_0\ell_0\propto 8\pi\eta^2$ (if assuming a global monople source). This line element exactly coincides with the one obtained in  \cite{Canate2016,Habib2012,tahamtan} for the same choice of $f(R)$ in (\ref{PFf}). 

The properties of the vacuum solution \eqref{exact1}--\eqref{exact2} with $\mu'=0$ given by the static metric \eqref{Schw} were discussed in detail in reference \cite{Canate2016}: it  exhibits an angle deficit $\Delta=\pi/2$ and a conical singularity with $R$ taking the same form as in \eqref{PF1} and can be associated with a global monopole, though the authors of \cite{Canate2016} interpret the constant $\mu_0\ell_0$ (denoted by ``$Q$'') as a geometric ``charge''  associated with the ``non--GR'' curvature terms. However, it is important to emphasise that the results of \cite{Canate2016} were obtained in the context of pure vacuum solutions, whereas we have obtained (\ref{Schw}) as the vacuum limit of the class of exact dust solutions \eqref{exact1}--\eqref{exact2}, and thus $\mu_0\ell_0$  can also be interpreted as a mass term associated with the global monopole source in an $f(R)$ geometry. In fact, as we show in the following section, when $\rho>0$ in a localised region the term $\mu_0\ell_0$ can be obtained from the same type of density integral (\ref{massInt0}) that defines Schwarzschild mass in GR in terms of the quasilocal Misner--Sharp mass integral. The suggested interpretation of $\mu_0\ell_0$ as a charge in \cite{Canate2016} perhaps follows from the term $\mu_0\ell_0/Y^2$ in \eqref{Schw} that is reminiscent of the electric charge in the a Riessner--Nordstr\"om metric. However, the appearance of a mass term with a quadratic power of $Y$ can also be understood as a byproduct of considering a different gravity theory from GR (irrespective of how physically plausible this theory may be). The causal and horizon structure of (\ref{Schw}) was discussed in detail in \cite{Canate2016}, so we remit the readers to this reference for further details. 

\section{Matching with a vacuum non--Schwarzschild solution.}\label{matching}

The Schwarzschild--like subcase discussed in the previous section can act as the vacuum exterior exterior to a localised non--infinitesimal spherical dust cloud given by \eqref{exact1}--\ref{exact2} with $\mu'\ne 0$. In other words: the class of dust solutions act as inner solutions or sources for the vacuum solution \eqref{Schw}. For this dust ball in an external vacuum situation to work we require the smooth matching along a common 3--dimensional interface given by $\Sigma=[t,r_b,\theta,\phi]$ for an arbitrary fixed $r=r_b$, with the inner spacetime being a dust cloud obtained from \eqref{exact1}--\ref{exact2} extending along $0\leq r\leq r_b$ and the exterior is the vacuum subcase $\mu'=0$ extending along $r\geq r_b$. 

It is known \cite{senovilla} that the conditions for a smooth matching of two spacetimes (``$\B_+$'' and ``$\B_-$'') along a common interface $\Sigma$ are more stringent in $f(R)$ gravity (when $f_{RRR}\ne 0$) than in GR. For the unit normal to $\Sigma$ given by $n_a=|Y'|/\sqrt{1-K}\delta_a^r$ (where the absolute value is meant to keep the same direction of $n_a$ under a sign change in $Y'$), the conditions for a smooth matching (without thin shells) are 
\ba \left[\hat h_{ab}\right]_\pm=0,\quad \left[K_{ab}\right]_\pm=0,\quad [R]_\pm=[R_{,a}]_\pm=0,\label{match}\ea 
where $[..]_\pm$ denotes the jump at $\Sigma$ as approached from $\B_+$ and $\B_-$, the Ricci scalar $R=1/Y^2+R_*$ is given by \eqref{PF1} and the intrinsic metric and extrinsic curvature of $\Sigma$ for the LTB metric are \cite{matravers,humphreys} 
\ba\fl \hat h_{ab}=g_{ab}-n_a n_b=\hbox{diag}[-1,0,Y^2,Y^2\sin^2\theta],\label{hab}\\
\fl K_{ab} = \hat h_a^c \hat h_b^d n_{c;d}=\hbox{diag}\left[0,0,\frac{|1-K|^{1/2}\,YY'}{|Y'|},\frac{|1-K|^{1/2}\,YY'}{|Y'|}\sin^2\theta\right],\label{Kab}\ea 
Notice that, besides continuity of the intrinsic metric and extrinsic curvature of $\Sigma$ (as in GR), the $f(R)$ geometry requires also the Ricci scalar and its gradients to be continuous at $\Sigma$ (see details in \cite{senovilla}). However, the fact that the LTB metric covers both spacetimes $\B_\pm$ in the same coordinate patch (admissible coordinates) is helpful to prove that there is sufficient parameter freedom in the solutions \eqref{exact1}--\ref{exact2} to fulfil conditions \eqref{match}. 

The solutions \eqref{exact1}--\ref{exact2} are fully determined by the free parameters $\lambda,\,\mu(r),\,\kappa(r)=K(r)-1/2,\,\tbb(r)$ (see \ref{append}), with the only parameter difference between the dust cloud $\B_-$ ($0<r<r_b$) and the Schwarzschild--like vacuum $\B_+$ ($r>r_b$) is the fact that $\mu$ must be a positive constant $\mu_b$ for $r>r_b$. Hence, the following are sufficient conditions for all of \eqref{match} to hold:
\ba\fl \mu(r_b)=\mu_b,\,\,\,\, \mu'(r_b)=0,\quad
[\kappa]_\pm=0,\,\,\,\, [\kappa']_\pm=0,\quad
[\tbb]_\pm=0,\,\,\,\, [\tbb']_\pm=0,\label{match2}
\ea
The proof is straightforward. Conditions $[\hat h_{ab}]_\pm=0$ and $[R]_\pm=0$ simply require the metric function $Y$ to be continuous, which follows readily if $\mu,\,\kappa,\,\tbb$ are themselves continuous ($C^0$) at $r=r_b$. Conditions $[K_{ab}]_\pm=0$ and $[R_{,a}]_\pm=0$ require also continuity of the gradients $\dot Y$ and $Y'$ at $r=r_b$, but the former is continuous (because of \eqref{exact2} and \eqref{quadr1}) if $Y$ is continuous, while $Y'$ is determined (see \eqref{Yprime}) by $\mu,\,\kappa,\,\tbb$ and the gradients $\mu',\,\kappa',\,\tbb'$,
 hence it is continuous if the free functions functions are $C^1$ at $r=r_b$. Notice that \eqref{match} only requires $\mu'(r_b)$ to vanish, but not the gradients $\kappa'(r_b)$ and $\tbb'(r_b)$. Also, the conditions for a smooth matching \eqref{match} and \eqref{match2} do not require $Y'(t,r_b)=0$.        

Given the smooth matching we discussed above, a Schwarzschild--like mass (related to the Schwarzschild mass $M_s$ of GR) follows from the integral (\ref{massInt}) as 
\begin{equation}\fl\frac{G}{c^2}
\int\limits_{0}^\pi\,\dd\phi\! \int\limits_{0}^\pi\dd\theta\,\sin^2\theta\!\int\limits_{0}^{r=r_b}{
    {\dd r\,\rho Y^2Y'}
             } =\frac{2\pi G}{c^2}\int_0^{r_b}{\rho Y^2Y'\dd r} =\frac{3\mu_b}{2},  \label{massInt0}\end{equation}
where we have neglected the monopole charge contribution, the angle deficit has been accounted for, $\mu_b=\mu(r_b)$ for an arbitrary fixed comoving radius $r=r_b>0$ and (again) we assume for simplicity that $\lambda=0$ in \eqref{exact1}. If $\mu(r)=\mu_b$ for all $r>r_b$ holds for any fixed $r_b$, we have in the range $0<r<r_b$ a dust ball around a global monopole with a conical singularity at $r=0$ with density (\ref{exact1}), matched to a vacuum spacetime \eqref{Schw} analogous to the Schwarzschild solution that occupies the range $r>r_b$. If $\lambda>0$ the same reasoning applies with an analogue of the Kottler (or Schwarzschild--de Sitter) metric.

It is worth noticing that the integral \eqref{massInt0} reinforces our claim that the term $\mu_0\ell_0$ in \eqref{Schw} should be interpreted as a mass term that can be generated by a source that is not (necessarily) a topological defect ({\it i.e.} dust), not as a geometric charge (as suggested in \cite{Canate2016}). Birkhoff's theorem holds, hence the preservation of the angular deficit at any $r=r_b$ is a necessary condition for the smoothness of this matching. These facts provide the sufficiency condition to prove that the dust spacetime that admits the local centre conditions \eqref{centre}--\eqref{centre2} must also exhibit the same angle deficit for every comoving domain.

\section{Regular dust wormholes without angle deficit.}\label{wormholes} 

To obtain fully regular dust solutions from (\ref{exact1})--(\ref{exact2}), without angle deficit and conic singularities, we need to remove the demand that centre conditions (\ref{centre}) and (\ref{centre2}) hold for finite curvature scalars. We also need to avoid dust configurations matched to the vacuum Schwarzschild--like spacetime \eqref{Schw} that follows from setting $\mu=\mu_b$ in a finite radial range. In other words, we need to choose the free functions $\mu,\,\kappa$ so that  $Y>0$ holds for all regular points $(t,r)$, with curvature singularities (Big Bang and Big Crunch) marked by points in the $(t,r)$ plane (with $r\ne$ constant) for which $Y(t,r)=0$ holds (for $r\ne$ constant) where $\rho,\,R,\,R_q$ diverge. The existence of LTB dust models without regular symmetry centres is a well known fact within the framework of standard GR (see \cite{kras2,hellaby,matravers,humphreys}).

In order to examine dust solutions (\ref{exact1})--(\ref{exact2}) without symmetry centres it is convenient to relabel the metric functions in \eqref{LTB} as
\begin{equation} Y(t,r) = a(t,r)\,F(r),\qquad K = k(r)\,F^2(r),\label{newvars}\end{equation}
where $a$ is a dimensionless scale factor and $F(r)>0$ holds for all $r$. The metric \eqref{LTB} and field equations (\ref{exact1})--(\ref{exact2}) take the forms
\ba \fl \dd s^2 = -\dd t^2 +a^2\left[\frac{\Gamma^2\,F'^2}{1-k\,F^2}\,\dd r^2+F^2\left(\dd\theta^2+\sin^2\theta\,\dd\phi^2\right)\right],\qquad \Gamma=1+\frac{a'/a}{F'/F},\label{LTB2}\\
8\pi\rho = \frac{\frac32\mu'}{a^3\Gamma\,F'}+\frac{\lambda}{2},\qquad 8\pi p=-\frac{\lambda}{2},\label{rho2}\\
\dot a^2 = \frac{\mu\,\ell_0}{a^2\,F^4}+\frac{1}{2F^2}-k+\frac{\lambda}{2}\,a^2,\label{dotYsq2}\ea
where now the radial coordinate range is no longer restricted to $r\geq 0$ (notice that $r$ lacks an invariant meaning, as \eqref{LTB2} is invariant under arbitrary rescalings $r=r(\bar r)$). 

The $f(R)$ dust solutions with an LTB metric that lack symmetry centres have analogous geometric properties as their LTB counterparts in GR, in particular their rest frames (3--dimensional hypersurfaces marked by constant $t$) are necessarily homeomorphic to either hyper--cylinders ${\bf S}^2\times {\bf{\textrm{R}}}$ (like the Schwarzschild--Kruskal manifold \cite{hellaby}) or 3--torii ${\bf S}^2\times {\bf S}^1$ (see  examples in \cite{matravers, humphreys}), both of which require $Y'=\Gamma\,F'=0$ to hold (for $\Gamma>0$) at a finite number of fixed values $r=r_*$ (``turning values'' where $Y'$ changes sign under regular conditions). 

The time evolution of these  exact LTB solutions lacking symmetry centres is given also by \eqref{sol1}--\eqref{sol8}. The effect of their ``unusual'' rest frames topology comes about in the choice of free functions (initial conditions) $F,\,\mu,\,k,\,\tbb$ subjected to the following regularity conditions:
\begin{itemize}
\item {\bf Avoidance of shell crossings.} Free parameters must be selected so that $\Gamma>0$ holds everywhere, as $\Gamma\to 0$ implies a shell crossing singularity with $\rho\to\infty$ for $Y>0$. However, the Big Bang and (for collapsing models) Big Crunch marked by $Y(t,r)=0$ for $r\ne$ constant are unavoidable.
\item {\bf Regularity at turning values $r=r_*$.} For $Y'=0$ to occur under regular conditions we must have $F'(r_*)=0$ with $\Gamma(t,r_*)>0$ for all $t$. From \eqref{LTB2}--\eqref{rho2} and the solutions \eqref{sol1}--\eqref{sol8}, this requires choosing the free functions $F,\,k,\,\mu,\,\tbb$ such that
\ba F',\,\mu',\,k',\,\tbb',\,1-k F^2\quad\hbox{have common zeros at all}\,\,r=r_*,\label{reg1}\\
 k(r_*)=\frac{1}{F^2(r_*)}>0.\label{reg2}\ea
\end{itemize}
Notice, in particular, that the regularity condition \eqref{reg2} rules out models obtained in \eqref{sol1}--\eqref{sol8} with $k\leq 0$ (equivalently $K\leq 0$ and $\kappa\leq 1/2$). In fact, as shown in \cite{bonnor} (see also \cite{matravers,humphreys}), the extrinsic curvature given by \eqref{Kab} for the 3--dimensional surfaces $\Sigma_*=[r=r_*,t,\theta,\phi]$ that mark the time evolution of the turning values of $Y'$ is discontinuous for LTB models of GR that do not comply with \eqref{reg2} (models with $k\leq 0$), thus requiring to apply the Israel--Lanczos formalism for a distributional source or thin layer at $r=r_*$ with surface density $\sigma$ and pressure $\pi=-\sigma/2$. The same problem occurs for the LTB solutions \eqref{sol1}--\eqref{sol8}. As a consequence, for $\lambda=0$ we have the same situation as in GR models with $\Lambda=0$ \cite{bonnor,suss3}: there are no fully regular ever--expanding solutions without symmetry centres and thus, only the re--collapsing solution \eqref{sol7} with $\kappa>1/2$ is free from surface layers at $r=r_*$. However, for $\lambda>0$ the condition $\kappa>1/2$ is not incompatible with ever--expanding solutions if the free functions are chosen so that quartic in \eqref{quadr1} satisfies $Q(Y)>0$ everywhere (see section 6 of \cite{suss3} for a detailed explanation of the analogous GR case).

To construct an LTB model with rest frames having a Schwarzschild--Kruskal ${\bf S}^2\times {\bf{\textrm{R}}}$ topology (as in \cite{hellaby}), we choose $F$ with a single turning value at (say) $r_*=0$, such that $F(0)$ is a global minimum: $F'(0)=0,\,F''(0)>0$, with $0<F(0)<F$ for all $r$ and $F\to\infty$ as $r\to\pm\infty$. We choose then the remaining free functions to comply with \eqref{reg1}--\eqref{reg2}, with equal signs for $F'$ and $\mu'$ (to keep $\rho>0$ in \eqref{rho2}), while $\Gamma>0$ requires the signs of $k',\,\tbb'$ to be (respectively) equal and opposite to that of $F'$. 

Models homeomorphic to ${\bf S}^2\times {\bf{\textrm{R}}}$ can also be constructed with an arbitrary number of turning values satisfying \eqref{reg1}--\eqref{reg2}, with free functions selected so that the signs of $F',\,\mu',\,k',\,\tbb'$ comply with $\rho>0$ and $\Gamma>0$. The simplest way to construct a model with toroidal topology ${\bf S}^2\times {\bf S}^1$ (see examples in \cite{matravers,humphreys}) is by setting up $n+m+1$ turning values, with one of them at $r_*=0$ and the remaining ones distributed as $m$ and $n$ in the ranges $r>0$ and $r<0$, so that $Y(t,r_m)=Y(t,r_n)$ for $r_m>0$ and $r_n<0$ marking the values of $r_*$ that lie furthest away from $r=0$ in each direction. The torus then follows by identifying the points of the 2--spheres marked by $r=r_m$ and $r=r_n$.    
            
\section{LTB solutions with nonzero pressure and critique of previous literature.}\label{pressure}

LTB solutions of $f(R)$ gravity with nonzero pressure follow from the ``inverse'' procedure described in section \ref{solving} in which the source is obtained after prescribing a metric (the LTB metric) and an ansatz for $f(R)$ such that $f_R$ is analytically invertible as $R=f^{-1}_R(Y)$ (similar procedures are discussed for FLRW metrics in \cite{Nojiri2011}). Ideally, it should be possible to find a choice of $f(R)$ that could yield a physically decent fluid, say a perfect fluid complying with a physically motivated equation of state. However, in practice it is extremely difficult to carry out this task without reverting to numerical methods. 

An evident first trial of this inverse procedure should be to consider the various $f(R)$ ansatzes in the literature that have so far fulfilled observational constraints: the proposals by Starobinsky \cite{Starobinsky2007} and Hu--Sawicki \cite{Hu2007} (see summary of most favoured $f(R)$ ansatzes in \cite{Jaime2012,luisa1}). Unfortunately, $f_R$ is not analytically invertible for the Starobinsky ansatz and it is only invertible for restrictive forms of the Hu--Sawicki ansatz, though the proposed exponential and logarithmic forms of $f(R)$ summarised in \cite{Jaime2012,luisa1} yield invertible forms for $f_R$. Therefore, just to illustrate this procedure we examine the fluid source obtained from choosing the  logarithmic ansatz for $f(R)$ considered by Miranda {\it et al} \cite{miranda} 
\begin{equation}\fl f(R)=R-\alpha\,R_0\ln\left(1-\frac{R}{R_0}\right),\qquad
 f_R=1+\frac{\alpha}{1+R/R_0}=1+\frac{Y}{\ell_0},\label{flog}\end{equation} 
where $\alpha,\,R_0$ are constants and we used \eqref{f1} with $\ell'_0=0$ and $\beta_0=1$. As a consequence of \eqref{flog} we have
\ba \frac{R}{R_0}=1-\frac{\alpha\ell_0}{Y},\qquad \frac{R_q}{R_0}=1-\frac32\frac{\alpha\ell_0}{Y},\label{miranda1}\\
\dot Y^2 = -K+\frac{\mu\ell_0}{4Y^2}+\frac{R_0\,Y^2}{12}\left(1-\frac{2\alpha \ell_0}{Y}\right),\label{miranda2}\\
{\cal M}=\mu+\frac{R_0 \ell_0 Y^3}{12}\left(\frac{Y}{\ell_0}-\alpha\right),\label{miranda3}\ea 
which substituted into \eqref{eqrho2}--\eqref{eqPP2} yields the following source variables
\ba\fl 8\pi\rho &=&  R_0\left[\frac14-\frac{\alpha}{4}\left(1+\frac{4\ell_0}{3Y}+2\ln\,\frac{\ell_0}{Y}\right)\right]-\frac{2}{\ell_0 Y}+\frac{\frac32\mu'}{Y^2Y'}+\frac{\mu'\ell_0}{Y^3Y'}-\frac{\mu\ell_0}{Y^4},\label{rho3}\\ 
\fl 8\pi p &=& -R_0\left[\frac14-\frac{\alpha}{12}\left(1+\frac{2\ell_0}{3Y}+6\ln\,\frac{\ell_0}{Y}\right)\right]+\frac{4}{3\ell_0 Y}+\frac{\mu'\ell_0}{3Y^3Y'}-\frac{\mu\ell_0}{3Y^4},\label{p3}\\
\fl 8\pi P &=& -\frac{\alpha R_0}{12}\left(1+\frac{2\ell_0}{Y}\right)-\frac{1}{3\ell_0 Y}+\frac{\mu'\,\ell_0}{6Y^3Y'}-\frac{2\mu\ell_0}{3Y^4},\label{P3}\ea
which evidently correspond to a mathematical fluid with various pathological properties: $R,\,\rho,\,p,\,P\to -\infty$ as $Y\to 0$, either along a centre worldline (hence no regular centres are possible) or at the Big Bang or Big Crunch singularities. While the fluid is grossly unphysical for small $Y$, it does mimic an asymptotically de Sitter ``Lambda dominated'' fluid for large $Y$: in the Friedman--like equation \eqref{miranda2} ($\dot Y/Y \to \sqrt{R_0/12}$) and in the fluid variables \eqref{rho3}--\eqref{P3}:
\begin{equation}\fl \frac{8\pi}{3} \rho\to \frac{R_0}{12}\left[1-2\alpha\ln\,\left(\frac{\ell_0}{Y}\right)\right],\qquad p\to -\rho,\qquad  8\pi P\to -\frac{\alpha R_0}{12},\end{equation}
which is consistent with the motivation expressed in \cite{miranda} behind the choice of $f(R)$ in \eqref{flog}. 

Analytic forms for non--static solutions have been found in the literature \cite{goswami,chakrabarti,cembranos} also following the ``inverse'' approach in which the field equations determine the source through a metric that is chosen {\it a priori}. The authors of \cite{goswami} consider an LTB--like metric and the ansatz $f(R)= R+\alpha R^2$, obtaining through the field equations the variables of a fully general fluid (including a heat flux term), yet they make no attempt to place restrictions on these variables. The only exact solution they obtain corresponds to the particular case $K=0$ of \eqref{LTB}. In \cite{chakrabarti} and \cite{cembranos} the authors consider also power law forms for $f(R)$ together with a very simple particular case of the LTB metric \eqref{LTB} with separable metric functions: $Y=r\,B(t)$ and $X(r)=1/\sqrt{1-K}$. Evidently, this excessive simplification of the LTB metric facilitates the integration of the field equations, but as a consequence it yields extremely restrictive solutions, all of which are necessarily particular cases of the solutions that can be obtained following the procedure outlined in this paper for the fully general LTB metric and  generic choice of $f(R)$. 

\section{Conclusion}\label{final}

We have obtained (see section \ref{dustsols}) the first example (as far as we are aware) of a class of exact solutions in $f(R)$ gravity whose source is non--static dust with an effective cosmological constant (which can always be set to zero). This exact solution is given by the LTB metric for the choice $f(R)\propto \sqrt{R}$ given in \eqref{PFf}. The full analytic integration for all parameter cases has been obtained (see \ref{append}) and its geometric and physical properties were discussed in detail (sections \ref{deficit}, \ref{vacuum}, \ref{matching} and \ref{wormholes}).

While these LTB dust solutions do not admit a regular symmetry centre, they can describe localised dust clouds that exhibit an angle deficit $\Delta=1/2$ associated with a global monopole at the centre (section \ref{deficit}). The vacuum limit (section \ref{vacuum}) is the solution examined in detail by \cite{Canate2016,Habib2012,tahamtan}, which can also serve as exterior field that matches smoothly with an inner localised dust cloud (section \ref{matching}).   Fully regular LTB dust solutions readily emerge (section \ref{wormholes}) if the spacetime manifold lacks symmetry centres, leading to dust wormholes whose rest frames (hypersurfaces of constant $t$) are homeomorphic to hyper--cylinders ${\bf \textrm{S}}^2\times {\bf \textrm{R}}$ (like the Schwarzschild--Kruskal manifold \cite{hellaby}) or to 3--dimensional torii ${\bf \textrm{S}}^2\times {\bf \textrm{S}}^1$ \cite{matravers,humphreys}.         

The dust solutions emerge from looking at the compatibility between the LTB metric and $f(R)$ gravity for an energy momentum tensor given by a fluid with anisotropic pressure. We have shown (see sections \ref{sec:field-equations} and \ref{solving}) how the integration of the field equations for this metric leads to an ``inverse'' procedure in which the Ricci scalar itself furnishes the evolution equation for the fluid layers, with the fluid variables fully determined by the field equations through a generic choice of $f(R)$, only restricted (to obtain analytic solutions) by demanding that $f_R=\dd f/\dd R$ be invertible to yield $R$ as a function of the main metric function $Y$. The dust solutions were readily obtained by pursuing the simplest condition for a perfect fluid within this procedure. 

As we argued in section \ref{pressure} using as example the $f(R)$ ansatz of \cite{miranda}, fluid solutions obtained for other admissible forms of $f(R)$ yield mathematical fluids which may be physically meaningful only in specific asymptotic ranges.  The non--static fluid solutions discussed in the literature that were are aware of (\cite{goswami,chakrabarti,cembranos}) can be derived from the inverse procedure we presented here as particular cases with separable metric.

Since $f(R)$ gravity theories can provide a promising alternative to GR that can explain cosmic evolution without assuming ``dark'' sources, we believe that it is worthwhile extending the research we have presented here to undertake the study of more general solutions. It is likely that the inverse procedure we have developed in this paper will yield mathematical fluids, however the exact solutions associated with such fluids (if not utterly unphysical) can serve as useful toy models to explore along a non--perturbative approach the effects (local and cosmological) of $f(R)$ geometry. Evidently, we cannot rule out that decent fluids could result from a fortunate choice of $f(R)$ ansatz.  We shall report work along these lines in the future. 

\section*{Acknowledgments.}

R. A. S. acknowledges financial support from grants SEP-CONACYT 239639 and PAPIIT-DGAPA RR107015.  L. G. J. acknowledges CONACYT postdoctoral fellowship 236937 and the Cosmology Group at the University of Heidelberg.             

\begin{appendix}

\section{Exact solutions.}\label{append}

Analytic solutions of the Friedman--like equation \eqref{exact2} follow from integrating the following quadrature
\begin{equation}\fl t-\tbb(r) = \FF(Y,\mu,\kappa,\lambda)\equiv \int{\frac{Y\dd Y}{\sqrt{Q(Y)}}},\qquad Q=\mu\ell_0-\kappa\,Y^2+\lambda\,Y^4,\label{quadr1}\end{equation}
where $\tbb$ is an arbitrary function (the Big Bang time), $\kappa=K-\frac12$ and we assume $\lambda=R_*>0$. The gradient $Y'$ can be obtained directly from the solutions of \eqref{quadr1} below, but a useful general form follows by implicit derivation  
\begin{equation}Y' =\frac{Y}{Q(Y)}\left[\tbb'-\frac{\partial \FF}{\partial\mu}\mu'-\frac{\partial \FF}{\partial\kappa}\kappa'\right].\label{Yprime}\end{equation}
This expression is used in section \ref{matching}. We have the following cases to integrate the quadrature \eqref{quadr1}:
\subsection{$\lambda>0$}

%
The solution of (\ref{quadr1}) and the kinematic evolution of dust layers depends on the existence of (up to two) real positive roots of $Q(Y)$:
\begin{equation} Y_\pm = \frac{\sqrt{\kappa\pm\sqrt{\kappa^2-4\lambda\mu\ell_0}}}{\sqrt{2\lambda}}.\label{rootsQ}\end{equation}
The following cases emerge:
\begin{itemize}
\item {\bf Ever expanding layers}. For $\kappa<2\sqrt{\mu\ell_0\lambda}$ the quartic $Q$ has no real positive roots. The quadrature (\ref{quadr1}) is 
\begin{equation}  t-t_0 = \frac{\sqrt{Q}}{2\lambda}+\frac{\kappa}{4\lambda^{3/2}}\ln\left[\sqrt{Q}+\frac{2\lambda Y^2-\kappa}{\sqrt{\lambda}}\right],\label{sol1}\end{equation}
so that dust layers are ever expanding (or ever collapsing if we set $t\to -t$) from a Big Bang singularity corresponding to $Y=0$ and marked by $t=\tbb(r)$ and given by
\begin{equation} \tbb = t_0+\frac{\sqrt{\mu\ell_0}}{2\lambda}+\frac{\kappa}{4\lambda^{3/2}}\ln\left(\sqrt{\mu\ell_0}-\frac{\kappa}{\sqrt{\lambda}}\right),\label{tbb} \end{equation}
with $Y\to\infty$ as $t\to\infty$.
\item {\bf Coasting layers}. For $\kappa=2\sqrt{\mu\ell_0\lambda}$ we have $Q=(\sqrt{\mu\ell_0}-\sqrt{\lambda}Y^2)^2$, hence  
\begin{equation}\fl \mp(t-t_0) = \frac{\pm(\sqrt{\mu\ell_0}-\sqrt{\lambda}Y^2)}{2\lambda}+\frac{\sqrt{\mu\ell_0}}{2\lambda}\ln\left[\pm(\sqrt{\mu\ell_0}-\sqrt{\lambda}Y^2)\right],\label{sol2}\end{equation}
which corresponds to a ``coasting'' behaviour in which, either dust layers emerge from (collapse to) the Big Bang at $Y=0$ towards (from) the asymptotic value $Y=(\mu\ell_0/\lambda)^{1/4}$, or emerge (collapse to) this asymptotic value towards (from) infinity $Y\to\infty$.
\item {\bf Bouncing layers}. For $\kappa>2\sqrt{\mu\ell_0\lambda}$ we have the following two cases. Layers emerge from the Big Bang at $Y=0$ reaching towards a maximal value $Y=Y_{-}$ given by (\ref{rootsQ}), bounce and collapse to a Big Crunch $Y=0$. The quadrature (\ref{quadr1}) has two branches:
\begin{equation}\fl \sqrt{\lambda}(t-t_0)= \left\{ \begin{array}{l}
 F(0,Y_\pm)-F(Y,Y_\pm),\qquad\hbox{expanding phase)}\,\,\dot Y>0 \\  
F(0,Y_\pm)+F(Y,Y_\pm),\qquad\hbox{collapsing phase)}\,\,\dot Y<0  \\ 
 \end{array} \right.\label{bounce1} \label{sol3}\end{equation}
where $0<Y<Y_{-}$ and
\begin{equation} F(Y,Y_\pm) = \frac12\hbox{arctanh}\left[\frac{Y_{+}^2+Y_{-}^2-2Y^2}{2\sqrt{Y_{+}^2-Y^2}\sqrt{Y_{-}^2-Y^2}}\right],\label{sol4}\end{equation} 
so that Big Bang time is $t=\tbb(r)$, while maximal expansion and collapse times are $\tmax=\tbb+F(0,Y_\pm)/\sqrt{\lambda}$ and $\tcoll=\tbb+2F(0,Y_\pm)/\sqrt{\lambda}$. 
In the other case layers collapse from $Y\to\infty$ towards a minimal value $Y=Y_{+}$ given by (\ref{rootsQ}) and bounce back to infinity. The quadrature (\ref{quadr1}) becomes:
\begin{equation}\fl \pm\sqrt{\lambda}(t-T)=\frac12\ln\left[2Y^2-Y_{+}^2-Y_{-}^2+2\sqrt{Q}\right]-\frac12\ln\left[Y_{+}^2-Y_{-}^2\right],\label{sol5} \end{equation}
where $Y>Y_{+}$ and the plus/minus sign respectively correspond to the expanding collapsing phases. 
\end{itemize}
%

\subsection{$\lambda=0$} 

The solutions of (\ref{quadr1}) are
\ba  Y = \sqrt{2}(\mu\ell_0)^{1/4}\sqrt{t-\tbb},\qquad\qquad\qquad \kappa = 0,\label{sol6}\\
     Y = \sqrt{t-\tbb}\sqrt{2\sqrt{\mu\ell_0}+|\kappa|(t-\tbb)},\qquad \kappa< 0,\label{sol7}\\
     Y = \sqrt{t-\tbb}\sqrt{2\sqrt{\mu\ell_0}-\kappa(t-\tbb)},\qquad \kappa> 0,\label{sol8}
    \ea
where $t=\tbb(r)$ marks the Big Bang time. As with the LTB dust solutions of GR with $\Lambda=0$, the cases $\kappa\leq 0$ correspond to ever expanding layers (where now $Y$ scales as $t^{1/2}$ and $t^{1/4}$ for zero and negative $\kappa$ (as opposed to $t^{2/3}$ and $t+\ln t$ in GR). Layers expand bounce and collapse to a Big Crunch when $\kappa>0$, reaching maximal expansion $Y=\sqrt{\mu\ell_0}/\kappa$ at $\tmax=\tbb+\sqrt{\mu\ell_0}/\kappa$ and collapse at $\tcoll=\tbb+2\sqrt{\mu\ell_0}/\kappa$.   

\end{appendix}

\end{document}